\documentclass[twocolumn,preprintnumbers,amsmath,amssymb,superscriptaddress]{revtex4}%
\usepackage{graphicx}
\usepackage{dcolumn}
\usepackage{bm}
\usepackage{amsmath}
\usepackage{amsfonts}
\usepackage{amssymb}
\usepackage{color}%

\usepackage[amssymb]{SIunits}
\usepackage{mathrsfs}
\usepackage{subfigure}
\usepackage{color}

\usepackage{graphicx}
\usepackage{dcolumn}
\usepackage{bm}
\usepackage{amsmath}
\usepackage{amssymb}
\usepackage{latexsym}
\usepackage{epsfig}
\usepackage{epstopdf}
\usepackage{amsbsy}
\usepackage{array}
\usepackage{amssymb}
\usepackage{bm}
\usepackage{float}
\usepackage{verbatim,color}

\begin{document}
\title{A controllable superconducting electromechanical oscillator with a suspended membrane}

\author{Yong-Chao Li}
\thanks{These authors contributed equally to this work.}
\affiliation{Research Institute of Superconductor Electronics, School of Electronic\\
Science and Engineering, Nanjing University, Nanjing 210093, China,}

\author{Jiang-shan Tang}
\thanks{These authors contributed equally to this work.}
\affiliation{School of Physics, Nanjing University, Nanjing 210093, China}
\affiliation{National Laboratory of Solid State Microstructures, College of Engineering and Applied Sciences, Nanjing University, Nanjing 210093, China}

\author{Jun-Liang Jiang}
\thanks{These authors contributed equally to this work.}
\affiliation{Research Institute of Superconductor Electronics, School of Electronic\\
Science and Engineering, Nanjing University, Nanjing 210093, China,}

\author{Jia-Zheng Pan}\affiliation{Research Institute of Superconductor Electronics, School of Electronic\\
Science and Engineering, Nanjing University, Nanjing 210093, China,}
\author{Xin Dai}
\affiliation{Research Institute of Superconductor Electronics, School of Electronic\\
Science and Engineering, Nanjing University, Nanjing 210093, China,}
\author{Xing-Yu Wei}
\affiliation{Research Institute of Superconductor Electronics, School of Electronic\\
Science and Engineering, Nanjing University, Nanjing 210093, China,}
\author{Ya-Peng Lu}
\affiliation{Research Institute of Superconductor Electronics, School of Electronic\\
Science and Engineering, Nanjing University, Nanjing 210093, China,}
\author{Sheng Lu}
\affiliation{Research Institute of Superconductor Electronics, School of Electronic\\
Science and Engineering, Nanjing University, Nanjing 210093, China,}
\author{Xue-Cou Tu}
\affiliation{Research Institute of Superconductor Electronics, School of Electronic\\
Science and Engineering, Nanjing University, Nanjing 210093, China,}
\author{Hua-bing Wang}
\affiliation{Research Institute of Superconductor Electronics, School of Electronic\\
Science and Engineering, Nanjing University, Nanjing 210093, China,}
\author{Ke-yu Xia}
\email{keyu.xia@nju.edu.cn}
\affiliation{National Laboratory of Solid State Microstructures, College of Engineering and Applied Sciences, Nanjing University, Nanjing 210093, China}
\affiliation{Collaborative Innovation Center of Advanced Microstructures, Nanjing 210093, China}
\author{Guo-Zhu Sun}
\email{gzsun@nju.edu.cn}
\affiliation{Research Institute of Superconductor Electronics, School of Electronic\\
Science and Engineering, Nanjing University, Nanjing 210093, China,}
\author{Pei-Heng Wu}
\affiliation{Research Institute of Superconductor Electronics, School of Electronic\\
Science and Engineering, Nanjing University, Nanjing 210093, China,}

\begin{abstract}
 We fabricate a microscale electromechanical system, in which a suspended superconducting membrane, treated as a mechanical oscillator, capacitively couples to a superconducting microwave resonator. As the microwave driving power increases, nonmonotonic dependence of the resonance frequency of the mechanical oscillator on the driving power has been observed. We also demonstrate the optical switching of the resonance frequency of the mechanical oscillator. Theoretical models for qualitative understanding of our experimental observations are presented. Our experiment may pave the way for the application of a mechanical oscillator with its resonance frequency controlled by the electromagnetic and/or optical fields, such as a microwave-optical interface and a controllable element in a superqubit-mechanical oscillator hybrid system.
\end{abstract}
\maketitle

There are intensive efforts for interfacing the microwave and optical domains by using various systems including the opto-electro-mechanical systems \cite{PRL.108.153603,PRL.108.153604,NatPhys.9.712,NatPhys.10.321,NatPhysEOC}, solid-state spins in resonators \cite{PRL.113.063603,PRL.113.203601, PRA.91.042307}, molecules or spins coupled to a nanowaveguide \cite{PRL.118.140501,PRA.97.052315}, electro-optical material-based Whispering Gallery mode resonator \cite{Optica.3.597} or cold atoms \cite{PRL.120.09301}. Among these approaches, the opto-electro-mechanical system is of particular interest because it is efficient in signal conversion and can be integrated on a chip.

In the opto-electro-mechanical resonator, the core element is the mechanical oscillator, which has been attractive for its promising applications in both classical and quantum regimes \cite{REV.SCI.INSTRUM,
PhysicsReports511,RevModPhys.86.1391}. The features of mechanical oscillators have been engineered by using various materials and structures. One of the main characteristics of a mechanical oscillator is its resonance frequency, which is determined by its parameters such as the mass, the geometric shapes and dimensions and the spring constant. These parameters are almost unchangeable after the sample is fabricated. In order to control the resonance frequency, an electrostatic field is usually applied to \emph{in situ} change the tension of the mechanical oscillator \cite{Appl.Phys.Lett.88.253101,Appl.Phys.Lett.107.113108,Appl.Phys.Lett.111.223108}. However adding the controlling component such as the electrostatic electrode to the structure of the mechanical oscillator greatly improves the complexity and difficulty of the whole micro-fabrication process.

In this letter we report an experimental implementation of controlling the resonance frequency of a mechanical oscillator with microwave and optical fields instead of the electrostatic field. We fabricate a microscale electromechanical system composed of a microwave coplanar waveguide (CPW) resonator and a mechanical oscillator, which is a suspended membrane. Under a microwave driving to the microwave subsystem, we observe a nonlinear and nonmonotonic frequency shift of the mechanical subsystem due to the radiation pressure force of the field. Also we use a laser beam pressing the mechanical part to modify its resonance frequency. Because the microwave signal passing through the microwave resonator is modulated by the motion of the mechanical oscillator, we demonstrate the switching on/off of the transmission of the sidebands of the microwave signal, which is a step towards bridging the microwave and optical signals without an optical cavity. The experimental observations are qualitatively reproduced with our theoretical models.

\begin{figure}
  \centering
 \includegraphics[width=0.9\linewidth]{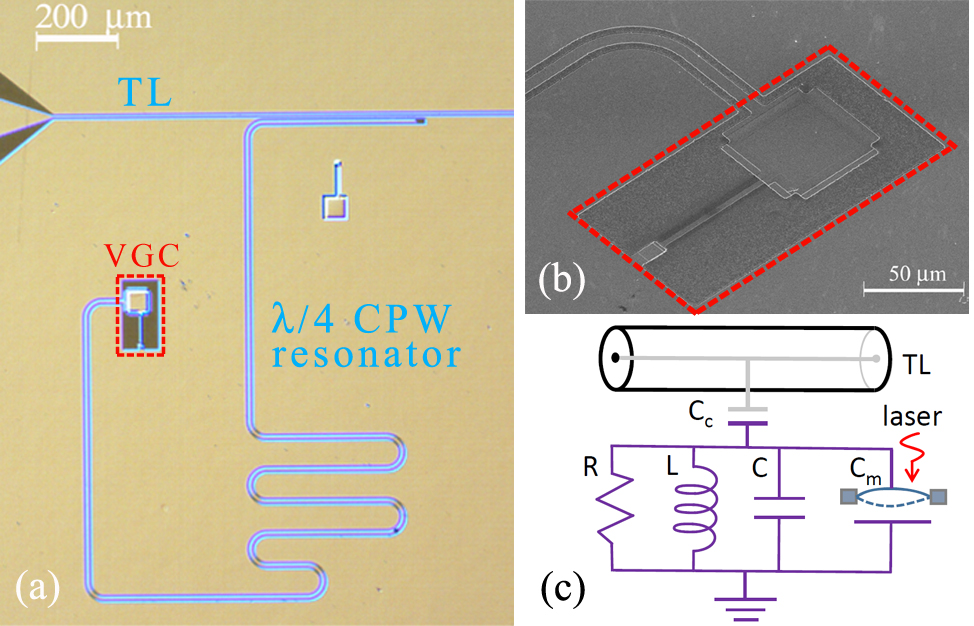}\\
  \caption{(\textmd{a}) Optical photograph of the sample, composed of a transmission line (TL), a $\lambda/4$ microwave coplanar waveguide (CPW) resonator, and a suspended micromembrane (marked by a red rectangular), fabricated on the end of the $\lambda/4$ CPW resonator. (b) SEM image of the VGC. The superconducting membrane is suspended from the substrate, forming a mechanical oscillator, and capacitively couples to the microwave CPW resonator. (c) Equivalent circuit of the device. The mechanical oscillator couples to the microwave CPW resonator via a capacitor $C_m$. The microwave decay is modeled by a resistor R. A laser beam (red arrow) can be applied to the mechanical oscillator to control its resonant frequency.}\label{fig:sample}
\end{figure}

The optical photograph of our sample is shown in Fig.~\ref{fig:sample}(\textmd{a}). A $\lambda$/4 microwave CPW resonator is capacitively coupled to a transmission line on one end and is terminated by the a vacuum-gap capacitor (VGC) \cite{doi:10.1063/1.3304168,nature2011NIST,Science.344.1262} (see Fig.~\ref{fig:sample}(b)) on the other end. The sample is made from aluminum on the high-resistivity silicon substrate evaporated in a multi-chamber evaporation system with ultra-high vacuum. The main fabrication procedure includes three steps: (1) Fabricating the lower electrode plate of VGC by lift-off process. (2) Preparing the sacrificial layer by using diluted S1813 UV photoresist. (3) Forming the rest of the circuit, including the top electrode plate of VGC, the $\lambda/4$ CPW resonator and the CPW feed lines by wet-etching after depositing another layer of $270~\nano\meter$-thick aluminum film. The fabricated device is equivalent to an electric circuit with a mechanical oscillator shown in Fig.~\ref{fig:sample}(c). The CPW resonator couples to the transmission line with characteristic impedance of  $50~\Omega$~via a capacitor of $C_c=4.8$~\femto\farad. The value of VGC is determined approximately by $C_m = \varepsilon_0 A/d$, where $\varepsilon_0$ is the permittivity of vacuum, $A=45 \times 45~\micro\meter^2$ is the membrane area and $d$ is the vacuum gap between the top and lower plates of VGC. The gap $d$ can be adjusted in the process of fabrication and later be tuned with the microwave and optical field. The displacement of the fundamental vibration mode of the square membrane is a spatial function given by
\begin{equation} \label{eq:z}
z(x,y) = z_0 sin(\pi x/L_x)sin(\pi y/L_y) \,,
\end{equation}
where $L_x$ and $L_y$ are the length and width of the suspended membrane, respectively, and the spatial ranges of vibration are $x \in [0, L_x]$ and $y \in [0, L_y]$, and $z_0$ is the oscillation amplitude. The fundamental-mode frequency is evaluated as $\omega_{11}=\sqrt 2 \pi P/\rho$, where $P$ and $\rho$ are the tension per unit length and mass per unit area, respectively.

\begin{figure}
  \centering
 \includegraphics[width=0.4\linewidth]{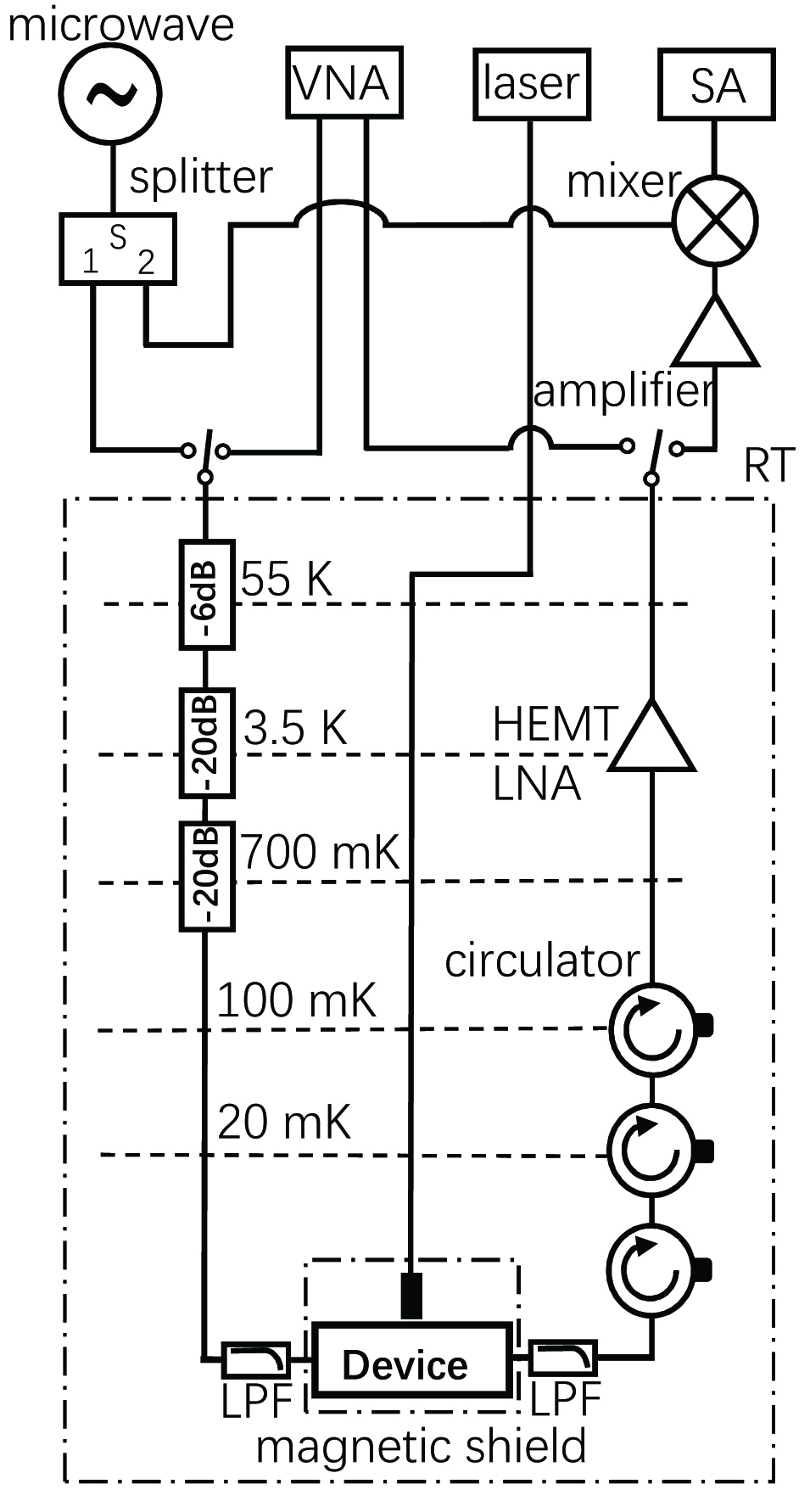}\\
  \caption{Measurement setup. Experiments are performed below $20~\milli\kelvin$ in an Oxford cryogen-free dilution refrigerator. A microwave signal near the resonance frequency of the microwave CPW resonator is applied to the electromechanical device through coaxial lines. Cryogenic attenuators ($20$~\deci\bel~ or $6$~\deci\bel) mounted in the input channel and the microwave circulators in the output channel are used to reduce noises. Both the input and output signals are filtered by low-pass filters (LPFs). The output signal from the device is first amplified by a low-noise amplifier (LNA), made of a high-electron-mobility transistor (HEMT), and microwave amplifiers at room temperature (RT). Then it is demodulated  by a mixer and read out by a spectrum analyzer (SA), from which the knowledge of the mechanical motion can be found.}\label{fig:measurement}
\end{figure}

The device is located in an Oxford Triton 400 dilution refrigerator below $20$~\milli\kelvin, with magnetic shielding at both mK and room temperature (see Fig.~\ref{fig:measurement}). To suppress the background noise from the higher-temperature parts, the microwave field input to the device is heavily attenuated at each stage of the dilution refrigerator and filtered by low-pass filters with a cutoff frequency of $12~\giga\hertz$. The output signal from the microwave resonator first passes through microwave circulators at cryogenic temperature to reject the back-action noise from amplifiers. Then, it is amplified by a low-noise amplifier and microwave amplifiers at room temperature, respectively \cite{PhysRevB.96.174518}. The low-noise amplifier uses a high-electron-mobility transistor located in the dilution refrigerator. A vector network analyzer is used to measure the transmission characteristics of the device. To measure the resonance frequency of the mechanical oscillator, we use frequency down-conversion technology as described below: The input microwave signal is divided into two paths. One of the paths is fed into the cryostat and the other one is for the local reference signal of a mixer. This mixer works as a frequency down-converter for the amplified output microwave signal which is modulated by the mechanical oscillator. The down-converted signal is then measured by a spectrum analyzer. We apply a $1310$~\nano\meter~ laser, generated by a semiconductor diode laser source, to the mechanical oscillator to control its resonance frequency, thus switch on/off the microwave sideband signals arising from the modulation of the mechanical oscillator.

\begin{figure}
  \centering
  \includegraphics[width=0.9\linewidth]{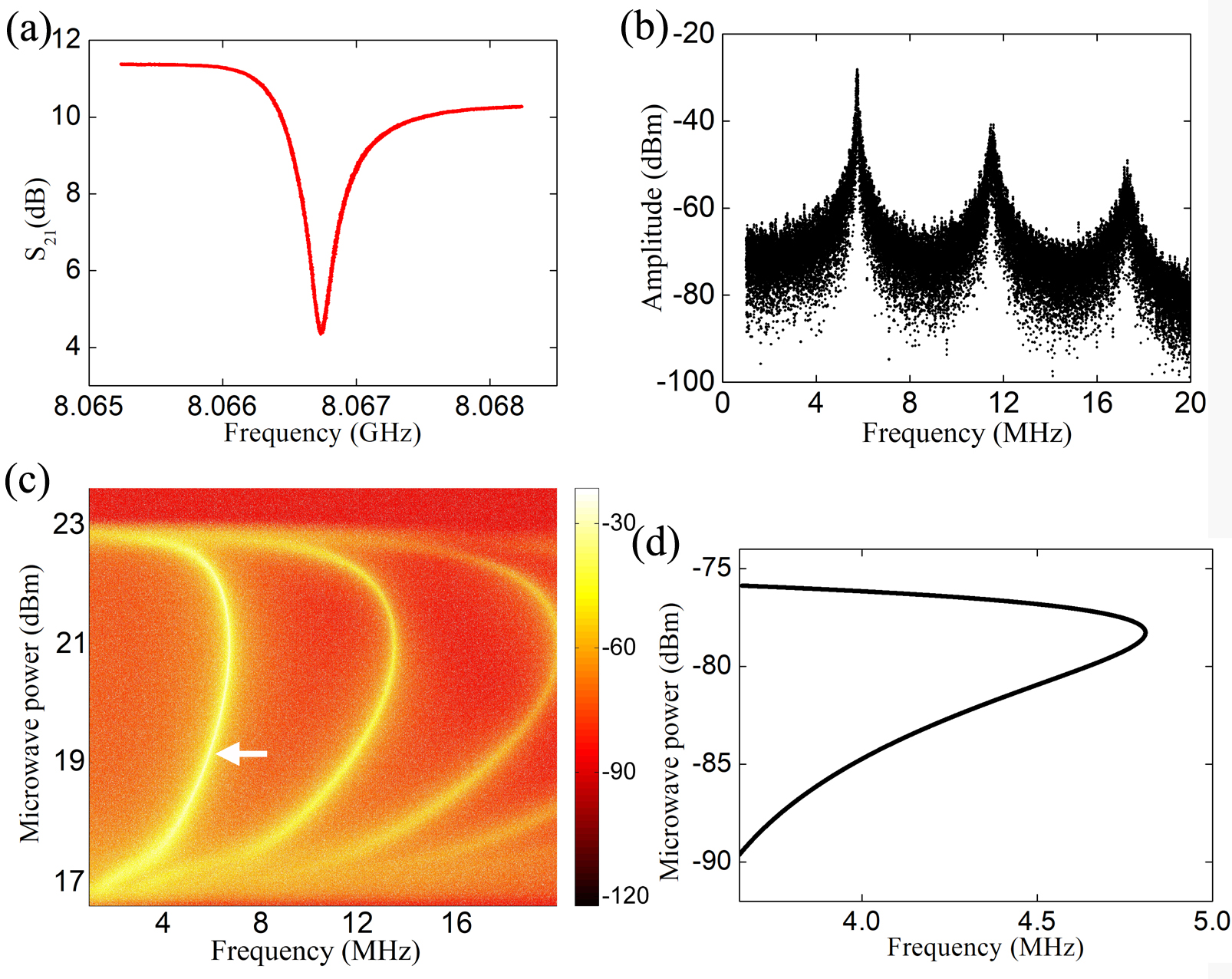}\\
\caption{(\textmd{a}) Measured $S_{21}$ of the transmission line. A resonance dip is obtained due to the coupling of the superconducting microwave resonant circuit to the transmission line. The resonant frequency and quality factor are 8.06674 GHz and $\sim 29000$, respectively. (b) Measured down-converted resonant frequencies with values of 5.76 MHz, 11.52 MHz and 17.28 MHz, respectively, corresponding to the input microwave power 19 dBm at room temperature. The quality factor of the mechanical resonator is about 670. (c) Dependence of mechanical fundamental-mode frequency and its harmonics on the input microwave power. The white arrow indicates the microwave power of 19.2 dBm where the laser is applied to the mechanical oscillator to control its resonant frequency. (d) Numerically calculated fundamental-mode frequency of the mechanical oscillator versus the input microwave power. Here we take parameters $\omega_c/2\pi = 8.06674$~\giga\hertz, $\omega_m/2\pi = 0.5$~\mega\hertz, $\kappa/2\pi = 305~\kilo\hertz$, $\Delta=0$, $g_0/2\pi = - 80.7$~\mega\hertz/\nano\meter~ and $m_\text{eff} = 100$~\pico\gram.} \label{fig:mw}
\end{figure}

The transmission characteristic $S_{21}$ of the CPW resonator is measured by a vector network analyzer shown in Fig.~\ref{fig:mw}(a). Clearly, a dip appears at the resonance
frequency of $8.06674~\giga\hertz$. In our device, the impedance of the CPW resonator is given by
\begin{equation} \label{eq:Zall}
{Z_\text{all}} = \frac{{ - j}}{{\omega {C_c}}} + {Z_0}\frac{{{Z_l} + {Z_0}\tanh \gamma s}}{{{Z_0} + {Z_l}\tanh \gamma s}} \;,
\end{equation}
 where ${Z_l} = 1/{j\omega {C_m}}$ is the VGC impedance, ${Z_0} = 50~ \Omega$ is the characteristic impedance of the CPW resonator and $\gamma$ is the propagation constant of microwave field in the vicinity of the resonance frequency. Under the superconducting condition and neglecting the loss, the propagation constant $\gamma$ approximates to the phase constant. $s$ is the physical length of the $\lambda/4$ CPW resonator. Using $Imag(Z_\text{all})=0$, $s=4500$~\micro\meter~and $\gamma=428.8$~\radian/\meter, we obtain $C_m = 1.0$~\pico\farad, in consistence with the estimate from the fabrication parameters.

After obtaining the resonant frequency of the superconducting microwave CPW resonator, we measure the frequency of the mechanical oscillator using the frequency down-conversion technology as described previously. As shown in Fig.~\ref{fig:mw}(b), with an input microwave power of 19 dBm at room temperature, three evenly spaced peaks appear at  $5.76$~\mega\hertz, $11.52$~\mega\hertz~ and $17.28$~\mega\hertz, respectively. Note that the lowest frequency $5.76$~\mega\hertz~ is the resonance frequency of the mechanical fundamental mode. The higher frequencies correspond to the higher-order harmonics of the fundamental mode. These harmonics are caused by the nonlinear conversion of the microwave resonator system \cite{ref18}. The dependence of the fundamental-mode frequency and its harmonics on the input microwave power are obtained by scanning the input microwave power. Hereafter we focus on the fundamental-mode frequency. It can be seen from Fig.~\ref{fig:mw}(c) that the fundamental-mode frequency of mechanical oscillator increases first from about $1$~\mega\hertz~ to a maximum $6.7$ MHz and then decreases rapidly as the input power increases further.

The observed nonmonotonic frequency shift of the mechanical oscillator can be understood by treating the device as an electromechanical resonator. Its motion is governed by the Hamiltonian \cite{ref19}
\begin{subequations} \label{eq:Hsys}
\begin{align}
 H & = H_m + H_\text{mw} + H_\text{in} \;,\\
 H_m & =  \frac{{{{ p}^2}}}{{2{m_\text{eff}}}} + \frac{1}{2}{m_{eff}}{\omega _m}^2{ z^2}  \;,\\
 H_\text{mw} & = - \hbar \Delta { a^\dag } a
  + i\hbar \sqrt {2{\kappa_e}} ({\alpha _\text{in}}{ a^\dag } - {\alpha _{in}}^* a) \;, \\
  H_\text{in} & =  \hbar {g_0} z{ a^\dag } a \;,
\end{align}
\end{subequations}
where $\omega_c$ is the resonance frequency of the microwave resonator, $m_\text{eff}$ and $\omega_m$ are the effective mass and the resonance frequency of the mechanical oscillator, $p$ and $z$ are the mechanical momentum  and displacement operators,  $a$ and $a^\dag$ are the annihilation and creation operators of the microwave resonator. $a^\dag a$ is the corresponding photon-number operator. The input photon flux is determined by the driving amplitude as $|\alpha_\text{in}|^2=P_\text{in}/\hbar\omega_c$, related to the input power $P_\text{in}$. The microwave resonator is driven by a microwave field with frequency $\omega_\text{in}$, yielding a detuning $\Delta=\omega_\text{in}-\omega_c$. The single-photon coupling rate of the electromechanical oscillator is $g_0= \partial \omega_c/\partial z$. For our device, $g_0 \approx -\alpha \omega_c/d$, where $d \approx 20$~\nano\meter. When considering the effective displacement of the whole area of the membrane, we have $\alpha < 0.4$. In a realistic experiment, $\alpha$ can be even smaller. We use a rough estimate of $g_0/2\pi \sim -80$~\mega\hertz/\nano\meter. The microwave loss of the system includes two contributions: one due to the coupling to the input and output channel yielding a decay rate $\kappa_e$, the other from the intrinsic loss causing a decay rate $\kappa_i$. For simplicity, we consider the critical-coupling case, i.e.  $\kappa_e= \kappa_i$.

For a constant driving $\alpha_\text{in}$ leading to all time derivatives ($\dot{a}(t), \dot{p}(t), \dot{z}(t)$) vanishing small, we can find the stable solutions $a(t)=\bar{a}$ and $z(t)=\bar{z}$ that \cite{ref19}
\begin{subequations} \label{eq:derivative}
\begin{align}
\bar a = \frac{{\sqrt {2{\kappa_e}} {\alpha _{in}}}}{{ - i(\Delta  - {g_0}\bar z) + \frac{\kappa }{2}}} \;, \\
\bar z =  - \frac{{\hbar {g_0}}}{{{m_\text{eff}}{\omega _m}^2}}{\left| {\bar a} \right|^2} \;,
\end{align}
\end{subequations}
where $\kappa=\kappa_i+\kappa_e$. Due to the displacement, the detuning now becomes $\bar{\Delta}= \Delta -g_0\bar{z}$.
By calculating the effective mechanical susceptibility, we obtain the effective resonance frequency of the mechanical subsystem under the driving $\alpha_\text{in}$ as
\begin{equation} \label{eq:omgeff}
\omega_\text{eff} = \sqrt {\omega^2 _0 + \omega _m^2} \;,
\end{equation}
where
\begin{equation} \label{eq:omega0}
{\omega _0}^2 = \frac{{\hbar {g_0}^2{{\bar a}^2}}}{{{m_\text{eff}}}}\left[ \frac{{\bar \Delta  + \omega_m }}{{{{(\bar \Delta  + \omega_m )}^2} + \left(\frac{\kappa}{2}\right)^2 }} + \frac{{\bar \Delta  - \omega_m }}{{{{(\bar \Delta  - \omega_m )}^2} + \left(\frac{\kappa}{2}\right)^2}} \right ] \;.
\end{equation}
In our case, the driving is strong that the frequency change of the mechanical subsystem can be even larger than its free oscillation frequency $\omega_m$. As a result, the commonly applied fluctuation approximation, i.e. $\omega_0 \ll \omega_m$,  breaks. The effective frequency $\omega_\text{eff}$ can be found by numerically solving the joint Eq.~\ref{eq:derivative} and then substituting the solution into Eq.~\ref{eq:omega0}. In fact, when the input power increases, the intra-cavity photon number $|\alpha|^2$ increases, and so does the displacement $\bar{z}$. Thus, the value of $\bar{\Delta} + \omega_m$ decreases rapidly. For a certain input power, the frequency $\omega_\text{eff}$ reaches its maximum. Crossing this point, $\bar{\Delta} + \omega_m$ approaches zero. As a result, the effective frequency reduces to a small value. Thus the effective frequency has a nonmonotonic dependence on the input power as observed in our experiment, which is also confirmed by the theoretical result in Fig.~\ref{fig:mw}(d). It can be clearly seen from Fig.~\ref{fig:mw}(d) that, by treating the device as a simple electromechanical resonator and using $\omega_c/2\pi = 8.06674$~\giga\hertz, $\omega_m/2\pi = 0.5$~\mega\hertz, $\kappa/2\pi = 305~\kilo\hertz$, $\Delta=0$, $g_0/2\pi = - 80.7$~\mega\hertz/\nano\meter~ and $m_\text{eff} = 100$~\pico\gram, we reproduce qualitatively the nonmonotonic frequency shift of the mechanical oscillator as the driving power increases. Due to the difficulty in the calibration on the power entering the microwave resonator, the input power in our model is different from the experimental data which is a nominal value at room temperature.

\begin{figure}
  \centering
  \includegraphics[width=0.9\linewidth]{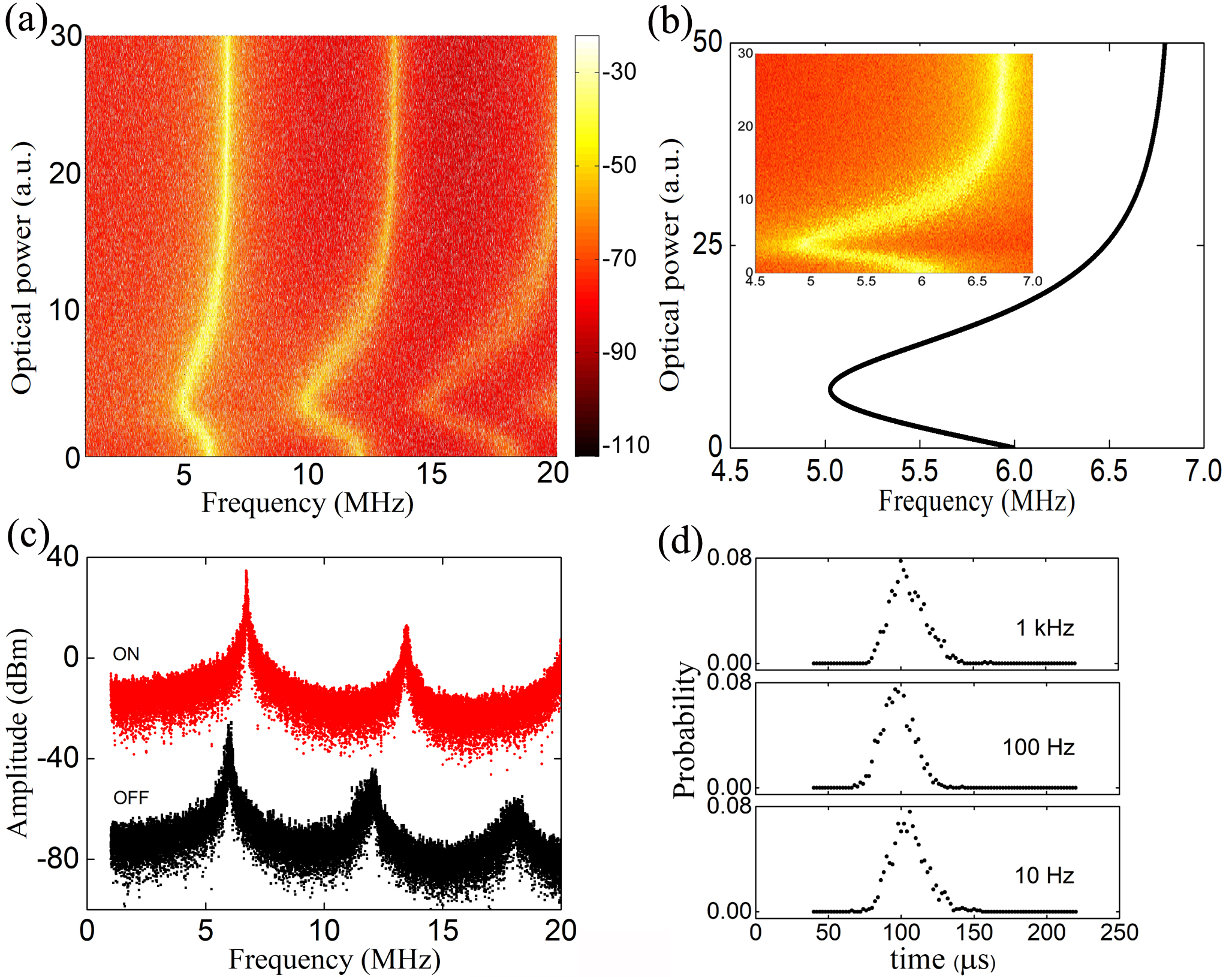}\\
  \caption{(\textmd{a}) Experimental dependence of the mechanical oscillator's fundamental-mode frequency and its harmonics on the optical intensity; (b) Theoretical dependence of the mechanical oscillator's fundamental-mode frequency on the optical intensity agrees qualitatively with the experimental data as shown in the inset, which is enlarged from (a). $\omega_c /2\pi = 8.06674$~\giga\hertz, $\omega_m/2\pi = 1.09$~\mega\hertz, $\kappa/2\pi = 1.614$~\mega\hertz, $g_0/2\pi = - 80.7$~\mega\hertz/\nano\meter,  $m_\text{eff}=100$~\pico\gram~ and $\Delta=0$. Due to the difficulty in the calibration of the optical power, we use the arbitrary unit in the y-axis of (a) and (b). (c) Measured mechanical resonant frequencies when the laser is on (red) and off (black). The data when the laser is on are shifted vertically for clarity. (d) The switching time statistic distribution of the down-converted microwave sideband signal with the mechanical fundamental-mode frequency of $6.71$~\mega\hertz~ switching from the one with a lower fundamental-mode frequency of $5.96$~\mega\hertz. The repetition rate of laser pulse is 10 Hz, 100 Hz and 1 kHz, respectively.} \label{fig:Laser}
\end{figure}

Then fixing the input microwave power at 19.2 dBm as schematically illustrated by the white arrow in Fig.~\ref{fig:mw}(c), we apply a $1310$~\nano\meter~ laser beam from a semiconductor diode to the suspended membrane. As shown in Fig.~\ref{fig:Laser}(a), we experimentally observe the laser-induced mechanical frequency shift. When the laser power increases, the mechanical fundamental-mode frequency deceases from $5.96$~\mega\hertz~ to $4.95$~\mega\hertz~ and then increase to $6.71$~\mega\hertz, wich also displays a nonmonotonic change. The first- and second-order harmonics show similar nonmonotonic behavior. Due to the difficulty in the calibration of the optical power, we use the arbitrary unit in the y-axis.

To understand this optically-induced frequency shift, we consider an optical pressure on the suspended membrane of the electromechanical system. Unlike typical optomechanical system \cite{ref19}, our model has an external force from the laser beam and the equation of motion of the mechanical oscillator with a microwave probe $\alpha_\text{in}$ takes the form
\begin{equation}
\ddot {z}(t) + {\Gamma _m}\dot {z}(t) + {\omega _m}^2 z(t) = \frac{{{F_{L}}}}{{{m_{eff}}}} - \hbar {g_0}\frac{{{{\left| {a(t)} \right|}^2}}}{{{m_{eff}}}} \;,
\end{equation}
where $F_{L} = 2\xi P_\text{L}/c$ is the optical force on the mechanical oscillator applied by the control laser beam with power $P_L$, $\xi$ is a coefficient describing how effective the laser pressure acts on the moving part,
and $\Gamma_m$ is the mechanical decay rate. Here we simply set $\xi=1$. $|\alpha|^2$ indicates the intra-cavity
photon number. $c$ is the vacuum light velocity. With this light force considered,
the stable solutions, $\bar{a}_L$ and $\bar{z}_L$, for the cavity mode $a$ and the mechanical
displacement $\bar{z}$ are determined by the joint equations
\begin{subequations} \label{eq:Oxa}
\begin{align}
{m_\text{eff}}\omega_m^2{{\bar z}_L} =  - \hbar {g_0}{\left| {{{\bar a}_L}} \right|^2} + {F_{L}} \; , \\
{{\bar a}_L} = \frac{{\sqrt {2{k_e}} {\alpha _\text{in}}}}{{ - i(\Delta  - {g_0}{{\bar z}_L}) + \kappa /2}} \;.
\end{align}
\end{subequations}%
Substituting the solutions into Eq.~\ref{eq:omgeff}, we obtain the effective mechanical resonance frequency $\omega_\text{eff}$ under the laser pressure. The force $F_\text{L}$ attributed to the laser beam is equivalent to the radiation pressure from the microwave CPW resonator. It also strongly modulates the mechanical resonance frequency. In this configuration, the mechanical oscillator vibrates initially at a resonance frequency, biased by the microwave input. When the laser beam is weak, the microwave radiation force is dominant. As the laser power increases, the light pressure presses the membrane and $\bar{z}_L$ becomes larger and larger. Thus, the effective detuning, $\bar{\Delta}=\Delta - g_0 \bar{z}_L$ increases. As a result, the intra-cavity photon number rapidly reduces, leading to smaller radiation force from the microwave field. When the laser beam is weak, the microwave radiation force is dominant and the total force on the membrane reduces. Thus the effective resonance frequency of the mechanical oscillator decreases rapidly. At a certain laser power, the frequency $\omega_m$ reaches its minimum. When the laser power increases further to become dominant over the microwave radiation force, the total force pressing the membrane increases again. Thus the effective mechanical resonance frequency increases but at a smaller rate. For a very strong laser beam, both the light pressure and the displacement $\bar{z}_L$ increases at constant rates, resulting in the saturation of $\omega_m$. Numerically solving Eq.~\ref{eq:Oxa}, we find the effective mechanical resonance frequency as a function of a constant laser power. As shown in Fig.~\ref{fig:Laser}(b), our theoretical model reproduces well the frequency shift of the mechanical oscillator taking $\omega_c /2\pi = 8.06674$~\giga\hertz, $\omega_m/2\pi = 1.09$~\mega\hertz, $\kappa/2\pi = 1.614$~\mega\hertz, $g_0/2\pi = - 80.7$~\mega\hertz/\nano\meter,  $m_\text{eff}=100$~\pico\gram~ and $\Delta=0$. $\omega_m$ and $\kappa$ change a little in comparison with those used in understanding the microwave-power-dependent frequency shift. One of the reasons may be the simultaneous injection of the laser and the higher-power microwave to the mechanical oscillator.

With the laser, we can optically control the sideband signals of the probe microwave field in our device.
The key idea is to apply a temporally modulated laser beam to the movable membrane and thus to change the capacitor $C_m$. In doing so, we dynamically modulate the resonance frequency of the microwave CPW resonator and subsequently its sideband signals. Experimentally, a rectangular voltage pulse is used to modulate the laser power. When the rectangular voltage pulse is on a high level, a laser beam is applied to press the suspended membrane. This pressure causes a frequency shift to the mechanical resonator, and thus the sideband signals of the resonant frequency in the microwave CPW resonator. The duration of such ``on''-state laser beam is controlled by the duty ration of the rectangular voltage pulse. The spectrum of down-converted microwave sideband signal is presented in Fig.~\ref{fig:Laser}(c). The fundamental-mode frequency shifts from $5.96$~\mega\hertz~ to $6.71$~\mega\hertz, when the laser is tuned from the ``off'' state to ``on'', corresponding to the laser power being 0 and $30$ as shown in Fig. 4(a), respectively. We can switch on/off the microwave sideband signal output from the microwave CPW resonator with an extinction ratio of $31~\deci\bel$.

Figure \ref{fig:Laser}(d) shows the switching temporal distribution of the transmitted ``on''-state sideband signal with the mechanical fundamental-mode frequency of $6.71$~\mega\hertz. This signal is switched from the ``off''-state signal, corresponding to the sideband field modulated by the lower mechanical fundamental-mode frequency of $5.96$~\mega\hertz. A time counter is triggered to start timing with the synchronization output of the pulse generator, which modulates the laser beam. The amplified down-converted signal is filtered with a bandpass filter through which only the signal associated with the mechanical fundamental-mode frequency of $6.71$~\mega\hertz~ can pass. This filtered signal is then sent to the time counter to stop timing. We perform $2000$ measurements for each repetition rate of the laser pulse to obtain the corresponding switching time statistic distribution. The highest repetition rate is at least $1$~\kilo\hertz, which is limited by the modulation rate of our laser. The average irradiation force of laser beam increases as the rectangular pulse repeats faster. If there is heating effect due to the laser, the mechanical resonant frequency will change \cite{PRL.101.197203}. In this case, the distribution will also change and even vanish as the repetition rate of laser pulse increases. However, as seen from Fig. 4(c)(d), little difference is observed in the distributions and the line width of the mechanical resonator when the repetition rate increases from $10$~\hertz~to $1$~\kilo\hertz. Therefore, the heat effect due to the laser is not observable with the measuring accuracy in our experiment.

To summarize, we fabricate a superconducting electromechanical system and modulate its mechanical frequency with both the microwave and optical fields. We first drive the microwave resonator strongly with the microwave field and observe large but nonmonotonic mechanical frequency shifts. By applying a laser beam to the mechanical membrane, we further demonstrate the optical control of the mechanical frequency and thus the transmission of the sideband signals of the microwave through the microwave CPW resonator. The developed device may work as an interface for microwave and optical domains. Also, the scenario demonstrated here has the potential in controlling a superconducting qubit coupled to a mechanical oscillator \cite{doi:10.1063/1.3304168,PRL.101.197203,Nature459,Nature494,PhysRevB.95.224515} with lasers mediated by an optically modulated capacitor.

This work was partially supported by the NKRDP of China (Grant Nos. 2016YFA0301801, 2017YFA0303703), NSFC (Grant Nos. 11474154,61521001,11874212,11574145), PAPD, and Dengfeng Project B of Nanjing University.


\newpage

\end{document}